\documentclass[twocolumn]{aastex631}

\newcommand{\ha}{${\rm H\alpha}$}

\begin{document}
\title{A Multi-Wavelength Investigation of Dust and Stellar Mass Distributions in Galaxies: Insights from High-Resolution JWST Imaging}

\correspondingauthor{Zhaoran Liu}
\email{zhaoran.liu@astr.tohoku.ac.jp}
\author[0009-0002-8965-1303]{Zhaoran Liu}
\affiliation{Astronomical Institute, Graduate School of Science, Tohoku University, 6–3 Aoba, Sendai 980-8578, Japan}

\author[0000-0002-8512-1404]{Takahiro Morishita}
\affiliation{IPAC, California Institute of Technology, MC 314-6, 1200 E. California Boulevard, Pasadena, CA 91125, USA}

\author[0000-0002-2993-1576]{Tadayuki Kodama}
\affiliation{Astronomical Institute, Graduate School of Science, Tohoku University, 6–3 Aoba, Sendai 980-8578, Japan}



\begin{abstract}
We study the morphological properties of mid-infrared selected galaxies at $1.0<z<1.7$ in the SMACS J0723.3-7327 cluster field, to investigate the mechanisms of galaxy mass assembly and structural formation at cosmic noon. 
We develop a new algorithm to decompose the dust and stellar components of individual galaxies by utilizing high-resolution images in the MIRI F770W and NIRCam F200W bands. Our analysis reveals that a significant number of galaxies with stellar masses between ${\rm 10^{9.5}<M_*/M_\odot<10^{10.5}}$ exhibit dust cores that are relatively more compact compared to their stellar cores. Specifically, within this mass range, the non-parametric method indicates that the dust cores are, on average, 1.23 ($\pm0.05$) times more compact than the stellar cores, when evaluated with flux concentration of the two components within a fixed radius. Similarly, the parametric method yields an average compactness ratio of 1.27 ($\pm0.06$). Notably, the most massive galaxy ($\rm{M_* \sim 10^{10.9}\,M_\odot}$) in our sample demonstrates a comparable level of compactness between its stellar core and dust, with a dust-to-stellar ratio of 0.86 (0.89) as derived from non-parametric (parametric) method. The observed compactness of the dust component is potentially attributed to the presence of a (rapidly growing) massive bulge, in some cases associated with elevated star formation. Expanding the sample size through a joint analysis of multiple Cycle~1 deep-imaging programs can help to confirm the inferred picture.
Our pilot study highlights that MIRI offers an efficient approach to studying the structural formation of galaxies from cosmic noon to the modern universe.
\end{abstract}



\section{Introduction} \label{sec:intro}
Galaxy formation and evolution is an active area of research in astrophysics that continues to present many unanswered questions and challenges. One of the most critical questions concerns the role of star formation in driving the growth of galaxies over cosmic time. Specifically, it has been of our particular interest to determine 
the primary mode of galaxy mass growth over the peak epoch of cosmic star formation \citep{Madau14}, during which many galaxies experienced intense star formation and formed the fundamental structural components (i.e., Hubble sequence). 
A critical insight into galaxies in this epoch has been obtained in the last two decades by the Hubble Space Telescope (HST) and adaptive optics assisted ground-based observations. With its high spatial-resolving power and near-infrared (NIR) sensitivity, HST has enabled us to glimpse the structural details of galaxy undergoing intense star formation activity and the formation of massive bulge \citep[e.g.,][]{dekel09, Conselice14}. 
However, studies have revealed that the structural evolution of galaxies over cosmic time is rather complex and heavily degenerated, and there are multiple pathways leading to the diverse structures and properties that we observe in the universe today. For example, one possible scenario for the most massive galaxy population is that they undergo gas ``compaction" and evolve massive bulges during cosmic time \citep{dekel14,Zolotov15, Lacerda20, Marques22}. The subsequent starburst and/or AGN feedback may then lead to quiescence \citep{genzel14, Yesuf14}, to galaxies of quenched dense core, e.g., those known as red nuggets \citep{Damjanov09, vanDokkum09, szomoru12, morishita&ichikawa16}. Through dry mergers, these structures eventually become the galaxies that we observe today, such as low-redshift spheroids and ellipticals \citep{barro13, vanderWel14, vanDokkum15,Tacchella16}. 

The study of the morphology of galaxies, encompassing the structure and shape, remains a crucial aspect in unraveling the unanswered questions in this field. Through a rigorous analysis of the morphology of galaxies, valuable insights into their formation history, such as the contribution of mergers, interactions, and other dynamical processes, can be gained \citep[e.g.,][]{Coziol07, Schreiber11, Cibinel19}.

A comprehensive understanding of the interstellar medium (ISM) is essential in unraveling the mechanisms that regulate star formation and how they differ across various regions of a galaxy \citep[e.g.,][]{abdurrouf22}.
One effective way to study the morphology of the ISM is through IR observations, as it can trace the emission from polycyclic aromatic hydrocarbons (PAHs) and dust continuum, which are closely related to star-forming regions \citep{wu05, brandl06, reddy06,Tielens08}. PAHs are organic molecules that are excited by ultraviolet (UV) radiation from young, massive stars, while dust continuum emission arises from thermal radiation emitted by dust particles heated by these same stars. The presence of PAHs and dust continuum emission can therefore be used as tracers of star formation activities within a galaxy. The {\it Spitzer Space Telescope} was the successor in the wavelength of interest \citep[e.g.,][]{smith07, lai20}; however, its primary exploration remained in nearby galaxies due to the limited resolution \citep[e.g.,][]{Boersma13, Stock13}.

Instead, a large part of this exploration has been taken placed by radio/submillimeter interferometers such as the Very Large Array (VLA) and the Atacama Large Millimeter/submillimeter Array (ALMA). Several pioneering studies have shed light on the formation of galaxy structure at cosmic noon \citep{simpson15, hodge16, barro16, tadaki17, tadakiapj17, elbaz18, Calistro18, fujimoto18,lang19, tadaki20, chen20, cheng20, guijarro22} and beyond \citep{Romano21, Pozzi21,Bowler22, fudamoto22}. However, studies of resolving gas and dust component are often limited to relatively bright sources, or have to come at significant expense of the facility's resource. 

With the launch of JWST \citep{Gardner06, Gardner2023}, the limitation has been largely lifted, opening a path for us to study ISM morphology of galaxies in unprecedented detail. Recently, morphological analysis of MIR emission in high-redshift galaxies has been enabled by high resolving power of MIRI \citep{cheng22, Inami22, shen23, Magnelli23}. 
In this work, we use JWST to study the spatially resolved ISM in galaxies, focusing specifically on the MIR band. Our goal is to address the difference between the mid-IR morphology and the stellar continuum morphology, with a particular emphasis on the compactness of the dust emission. By doing so, we aim to gain new insights into the mechanisms that govern star formation within galaxies.

Where relevant, we adopt the AB magnitude system \citep{oke83,Fukugita96}, cosmological parameters of $\Omega_m=0.3$, $\Omega_\Lambda=0.7$, $H_0=70$\,km\,s$^{-1}\,{\rm Mpc}^{-1}$, and the \citet{Chabrier03} initial mass function (IMF).

\section{Data and sample selection} \label{sec:style}
\subsection{SMACS J0723.3-7327: An optimal field to initiate a new study of galaxy morphology}
The sample examined in this study is selected from the background galaxies of the massive lensing cluster SMACS0723, which is located approximately 4.24 billion light-years away from us. The JWST observations used in this work were obtained as part of the Early Release Observations (ERO) program (ID 2736; PI: Pontoppidan, \citealt{Pontoppidan22}), and were released on July 14th,  2022, the high quality of the images and spectra of SMACS0723 have been obtained in particular with the instruments NIRCam and MIRI \citep{nircam1, miri1}. The data were retrieved from the Mikulski Archive for Space Telescopes (MAST) at the Space Telescope Science Institute, the specific observations analyzed can be accessed via \dataset[DOI: 10.17909/35k9-w010]{https://doi.org/10.17909/35k9-w010}. 

SMACS0723 was previously observed with HST as part of the Reionization Lensing Cluster Survey program \citep[RELICS;][]{2019ApJ...884...85C,2020ApJ...889..189S}. 
The cluster's mass distribution creates a strong gravitational potential well that bends and focuses light from objects located behind it, resulting in brighter and distorted images. 
The background galaxies along the sightline, even though much further away than the cluster itself, are magnified by the gravitational lensing effect, allowing the study of those distant galaxies with increased resolution and greater details. In addition, lensing magnification can help overcome the effects of point spread function (PSF) by magnifying the shape and brightness of distant galaxies, allowing us to study their true morphology at sub-resolution scales. This magnification effect acts as a natural zoom lens, providing greater accuracy in observing the inner regions of a galaxy, which is particularly useful for our work as we aim to precisely locate the star formation regions within galaxies. Various studies have utilized gravitational lensing effects to investigate high-redshift galaxies \citep{zheng12, coe13, livermore17, yang22}.

\subsection{NIRCam+HST Photometry and photmetric redshift}\label{sec:nircam}
For source identification, we start with a catalog generated in \citet{morishita22}. The work reduced the raw data of the available JWST and HST data in the field using {\tt borgpipe} \citep{morishita21}. The reduced data set also includes the NIRISS direct images, which were taken as part of the WFSS mode observations. The catalog comprises photometric fluxes of NIRCam (F090/150/200/277/356/444W), NIRISS (F115/200W), and HST WFC3-IR and ACS (F435/606/814/105/125/140/160W), which was retrieved from MAST in the form of seven filters. 
We also adopt the photometric redshift catalog derived in the same study, computed by using {\tt{EaZY}} software (\citealt{Brammer08}). 
For further details regarding the catalog, the readers are referred to \citet{morishita22}. 
\subsection{MIRI Photometry}\label{sec:miri}
Photometric fluxes of sources in MIRI images are obtained separately from the optical-NIR images aforementioned. We start from the fully-reduced MIRI images (F770/1000/1500/1800W) processed and made available by \texttt{Grizli}\footnote{\url{https://s3.amazonaws.com/grizli-v2/JwstMosaics/v4/index.html\#smacs}} \citep{brammer22_grizli} .

We align the MIRI images to the NIRCam F200W, and then refine the pixel grid of all images to match that of the F200W image using a Python package reproject.\footnote{\href{https://reproject.readthedocs.io/en/stable/}{https://reproject.readthedocs.io/en/stable/}}
The object detection is performed on a single image, the MIRI $7.7\,\mu$m image, by running {\tt{SExtractor}} \citep{bertin1996}. We implement the detection strategies described in \citet{iani22} and adopt their configuration, which is optimized to detect the fainter sources as shown in a previous survey \citep{2013ApJS..206...10G}. We apply the corresponding weight map to improve the rejection of spurious sources. We measure photometric fluxes of each source by adopting \citet{kron1980} aperture MAG\underline{\hspace{0.5em}}AUTO photometry.
We run {\tt{SExtractor}} individually on the remaining three bands with the same procedure, by repeating this procedure, we obtain four catalogs for F770/1000/1500/1800W, respectively. We then cross-match sources among these four catalogs with the matching radius set to $0.\!''5$ from the sources in the F770W detection. 
To ensure reliable photometry and morphology analysis afterwards, we select only those with MAG\underline{\hspace{0.5em}}AUTO brighter than the $3\,\sigma$ limiting magnitude and have a signal-to-noise ratio (SNR) $>$ 3 in F770W, which results in 442 galaxies.
For subsequent analysis, we adopt MAG\underline{\hspace{0.5em}}AUTO for our selected sources.

\subsection{Sample selection}\label{sec:sample selection}
We use the MIRI catalog generated in the previous section and select sources for further analysis as follows:
\begin{itemize}
\item Firstly, we perform a cross-match between the MIRI catalog and the NIRCam+HST catalog (Sec.~\ref{sec:nircam}) using the matching radius $0.\!''5$.
\item We then limit our sample to galaxies within the photometric redshift range of $1 < z < 1.67$, as this range corresponds to the wavelength sensitivity of the MIRI F770W filter to capture the 3.3 $\mu$m PAH emission feature. A strong feature of PAH, if present, can be observed as a flux excess from the adjacent filters that cover the continuum.
\item Lastly, we conduct spectral energy distribution (SED) fitting for the selected galaxies in the previous step. To ensure reliable results, we refine our sample by including only those with the SED fitting results with a reduced $\chi^2_\nu$ value of less than 3. 
We exclude those galaxies that do not have any dust emission inferred by the best-fit model (i.e. MIRI bands are primarily dominated by stellar components). 
\end{itemize}

Through the three steps above, we obtain 27 galaxies for the following analysis.

\section{Physical Properties of the Sample Galaxies} \label{sec:floats}
\subsection{Lens Model}\label{sec:lensmodel}
In order to correct the magnification effect by the foreground cluster and derive the intrinsic physical properties of our sample galaxies, we utilize the lensing model that was recently presented in the paper by \cite{2022ApJ...938...14G}. 
This HST-based lensing model was generated with the Light-Traces-Mass approach (LTM; \citealt{2005ApJ...621...53B, 2009MNRAS.396.1985Z}), which complements the parametric models already available in the RELICS repository and elsewhere and thus helps span a representative range of solutions. The magnification factor, denoted by $\mu$, is calculated using this model. The intrinsic flux is calculated by simply dividing the observed flux by this factor.
\begin{figure*}
\plotone{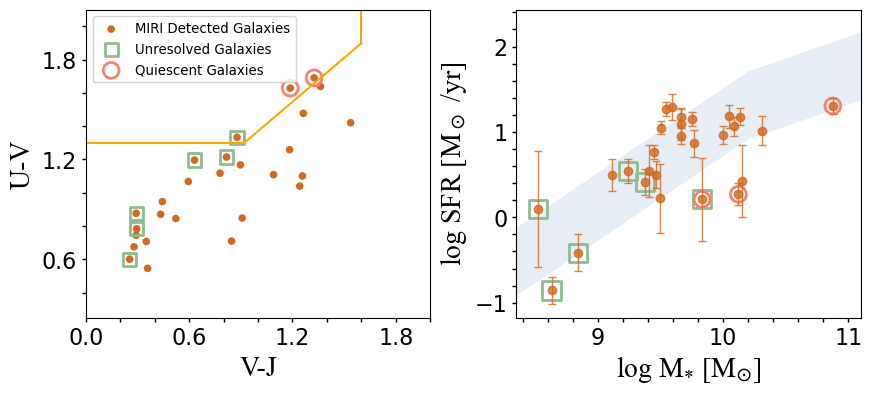}
\caption{(Left): Rest-frame $UVJ$ diagram used to classify galaxy populations. $U-V$ and $V-J$ colors are calculated by SED fitting. The orange solid lines represent the boundaries defined in \cite{williams2009}. Dark orange circles represent all MIRI selected sources, while green squares indicate unresolved galaxies that were excluded from the morphology analysis as explained in Sec.~\ref{sec:decomposition}, $UVJ$-selected quiescent galaxies are marked as red open circles.
(Right): Stellar mass vs. SFRs plot for our samples. The symbols used in this panel are the same as in the left panel. The error bars for SFR are computed based on {\tt{CIGALE}}'s Bayesian probability distribution.
 A blue-shaded region indicates the $\pm$0.4 dex range of the star formation main sequence at $1<z<1.5$ derived by 
\citet{2014ApJ...795..104W}.
\label{fig:main sequence and uvj}}
\end{figure*}

\subsection{Stellar Mass and Star Formation Rate}
\label{sec:stellarmass_sed}
\subsubsection{SED fitting}\label{subsec:sed}
We perform SED fitting using the broad-band photometry to infer stellar mass and dust extinction. We use Code Investigating GALaxy Emission ({\tt{CIGALE}}, \citealt{2019A&A...622A.103B}), which is optimized for data of wide range wavelength coverage, from X-ray to radio. The code is designed to self-consistently estimate the infrared dust reemission in infrared wavelengths for the absorbed UV energy by ISM. We adopt a delayed exponential star-formation history ({\tt sfhdelayed}) with the functional form of ${\rm SFR}(t) \propto t \exp({-t}/{\tau}$). 
The stellar population synthesis model is taken from \cite{2003MNRAS.344.1000B} with solar metallicity. Dust attenuation follows the \cite{Calzetti2000} extinction law. More details of our SED fitting parameter settings are shown in Table \ref{tab:CIGALE}.
\begin{table}
\centering
\caption{A list of parameters used for the CIGALE SED fitting.}
\label{tab:CIGALE}
\begin{tabular}{c|c} 
    \hline
    \hline
    Parameter name & Value \\
    \hline
    \multicolumn{2}{c}{sfhdelayed}\\
    \hline
    $\tau_{\text{main}}$[Myr] & 100, 300, 1000, 2000, 4000, 6000, 8000\\
    age[Myr] & 100, 500, 1000, 2000, 3000, 4000, 5000 \\
    $\tau_{\text{burst}}$[Myr] & 50\\
    $f_{\text{burst}}$ &  0.0, 0.01, 0.05, 0.1\\         
    \hline
    \multicolumn{2}{c}{SSP: \cite{2003MNRAS.344.1000B}}\\
    \hline
    IMF & \cite{Chabrier03} \\
    metallicity & 0.02 \\
    \hline
    \multicolumn{2}{c}{Dust attenuation: \cite{Calzetti2000}} \\
    \hline
    E(B-V)$_*$ & 0, 0.1, 0.2, 0.3, 0.4, 0.5, 0.6, 0.7\\
    & 0.8, 0.9, 1, 1.2, 1.5, 1.6, 1.7  \\
    E(B-V)$_\mathrm{factor}$ & 0.44 \\
    \hline
    \multicolumn{2}{c}{Nebular emission} \\
    \hline
    log U & -2.6, -2.7, -2.8, -2.9, -3.0, -3.1, -3.2   \\
     $Z_{\text{gas}}$ & 0.014 \\
     \hline
    \hline		   
\end{tabular}
\end{table}

\subsubsection{Rest-frame UVJ diagram and Star-Forming Main Sequence}
\label{sec:uvj_sfr}
The selection of the parent sample in this study is based on detection in the MIRI F770W, which is sensitive to the PAH emission associated with young, massive stars. Thus, the majority of the selected sources are expected to be star-forming. To diagnose their star-forming phase, we adopt the rest-frame $UVJ$ color-color diagram \citep{williams2009}. We adopt this method by utilizing the rest-frame colors of galaxies in the $U$, $V$, and $J$ filters calculated by using the best-fit SED template.
As displayed in Fig.~\ref{fig:main sequence and uvj}, three galaxies in our final sample are located in the quiescent region of the $UVJ$ diagram, but rather at the boundary between the quiescent and star-forming regions. 

In addition to the $UVJ$ classification, we also estimate the total SFR of our sample by integrating the unobscured and obscured star formation. We determine the rate of unobscured star formation from the rest-frame $2800$\,\AA~luminosities as they trace the UV emission from young, massive stars. We then estimate the rate of obscured star formation from the total IR luminosities. The $2800$\,\AA~luminosities and total IR luminosities are both based on best-fit SED model. We calculate the SFRs by applying the methodology proposed by 
\cite{Bell05}, employing the subsequent equation:
\begin{eqnarray}
\mathrm{SFR(M_\odot/yr)= 1.09 \times 10^{-10}} [L_\mathrm{{IR}} + 2.2L_{\mathrm{UV}}](\mathrm{L_\odot})
\end{eqnarray}
The prescription, originally proposed by \cite{kennicutt98} and subsequently revised by \cite{Bell05}, incorporates scaling factors based on the IMF prescribed
by \cite{Chabrier03}. As illustrated in Fig.~\ref{fig:main sequence and uvj}, we find that the majority of our sources are located on or near the star formation main sequence predicted by \cite{2014ApJ...795..104W}. We also identify several galaxies located below the main sequence, including the three $UVJ$-quiescent galaxies identified above.

\begin{figure*}
\plotone{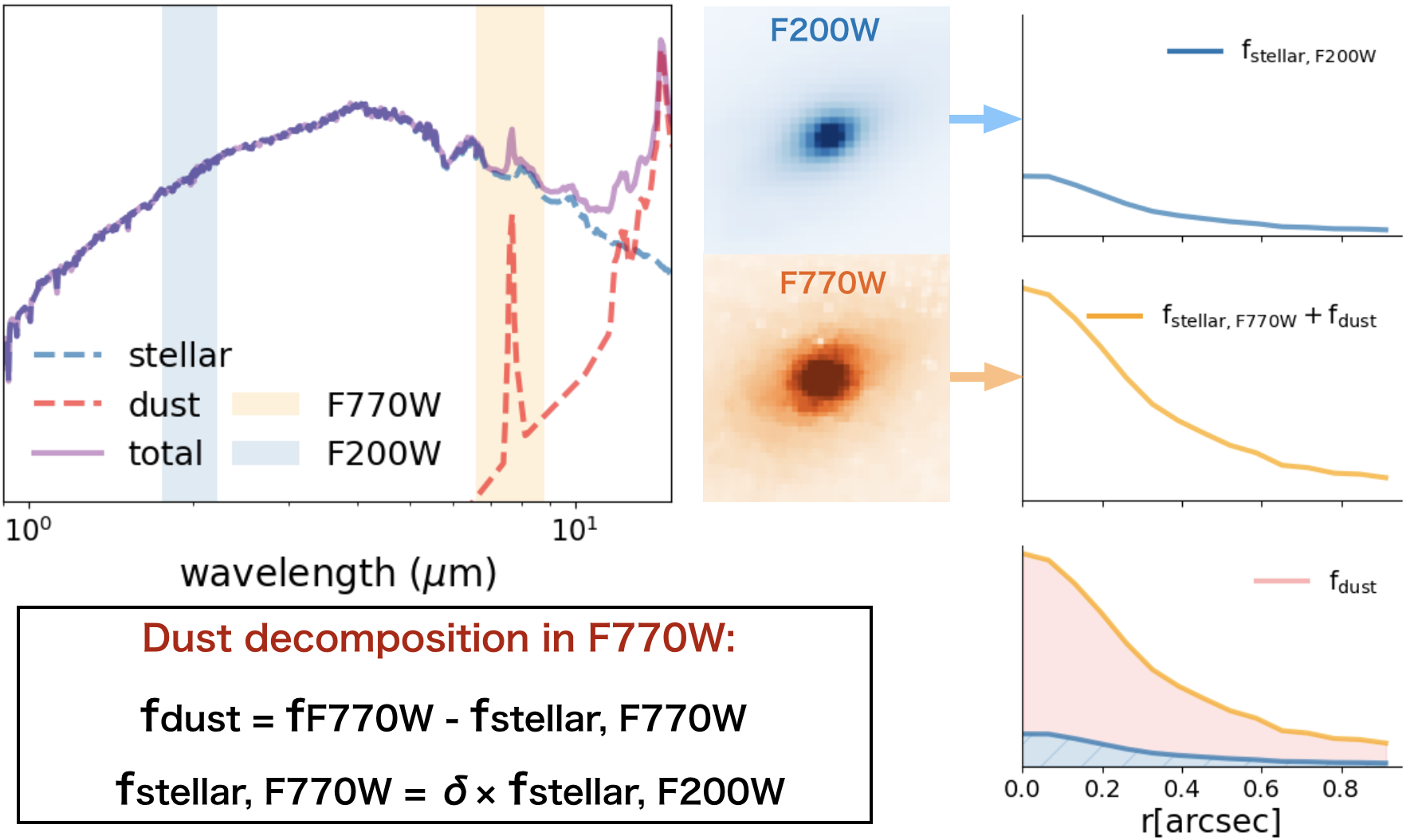}
\caption{A schematic view of our dust decomposition method.
(Top left): An example SED from our galaxy sample at redshift 1.32 features the 3.3 $\mu$m PAH emission captured by the F770W filter. (Bottom Left): Dust decomposition utilizing observed F770W and observed F200W, $\delta$ denotes stellar flux density ratio between two filters. (Right): From top to bottom, the flux profile of the F200W filter (representing the stellar continuum), the F770W filter (capturing a mixture of stellar continuum and dust emission), and the extracted dust emission.
\label{fig:examplesed}}
\end{figure*}

\subsection{Dust and Stellar Morphologies
}
\label{sec: 1d_profile_and_galfit}

As stated in Sec.~\ref{sec:intro}, PAH emission is known to be a reliable tracer of star-forming activities. However, the MIRI F770W band, which we use to cover the $3.3\,\mu$m feature in this study, may has a significant contribution from stellar continuum as well.
Thus, in the following subsection we present a new algorithm to decompose the two components in the one-dimensional radial profile. 
Our approach is visualized in a schematic in  Fig.~\ref{fig:examplesed}.
It is noted that, to assess any systematic differences that originates in one-dimensional profile measurements, we apply the following decomposition method by using both results from non-parametric (Section~\ref{sec: 1d model introduction}) and parametric  (Section~\ref{sec:parametric}) methods.



\subsubsection{Decomposition of Dust and Stellar Profiles}\label{sec:decomposition}

For the redshift range of this study, the one-dimensional radial profile in F200W, $f_{\mathrm{F200W}} (r)$, can be considered as the stellar profile, since it is dominated primarily by stellar continuum (see top left panel in Fig.~\ref{fig:examplesed}): 
\begin{equation}
f_{\mathrm{stellar}}(r) = f_{\mathrm{F200W}}(r)
\end{equation}

On the other hand, the observed F770W light profile, $f_{\mathrm{F770W}}(r)$, is the sum of the $3.3\,\mu$m PAH emission and the stellar continuum (primarily by low-mass stars; \citealt{Kajisawa15}) at the corresponding wavelength: 
\begin{equation}
f_{\rm F770W}(r) = f_{\mathrm{stellar,F770W}}(r) + f_{\rm dust, F770W}(r)
\end{equation}

To decompose the dust component in F770W, we perform the following two steps:
\begin{equation}
f_{\mathrm{stellar, F770W}}(r) = \delta_{\mathrm{200}}^{\mathrm{770}} \times f_{\mathrm{stellar, F200W}}(r)
\end{equation}
\begin{equation}
f_{\rm dust}(r) = f_{\rm F770W}(r) - f_{\mathrm{stellar,F770W}}(r)
\end{equation}
The quantity $\delta_{\mathrm{200}}^{\mathrm{770}}$ is the ratio of stellar continuum flux density between the F770W and F200W filters, calculated with the stellar component of the best-fit SED obtained in Sec. \ref{subsec:sed}. We note that the F200W image is convolved to the PSF size of F770W before deriving the one-dimensional profile.

Based on the method above, we obtain one-dimensional radial flux density profiles for both the stellar component and dust component, as well as their corresponding errors obtained from the rms map. 

Lastly, to quantify the compactness of individual galaxies, we define a flux-based compactness parameter for both dust ($c_{\rm dust}$) and stellar components ($c_{\rm stellar}$) as the ratio of the flux inside 1\,kpc to the total flux:
\begin{equation}\label{eq:6}
c_{\rm stellar} = \int^{\rm 1kpc} f_{\rm stellar}(r) rdr/ \int^{r_{\rm out}} f_{\rm stellar}(r) rdr
\end{equation}
\begin{equation}\label{eq:7}
c_{\rm dust} = \int^{\rm 1kpc} f_{\rm dust}(r) rdr/ \int^{r_{\rm out}} f_{\rm dust}(r) rdr
\end{equation}

Due to the varying sensitivities of the F200W and F770W filters, the SNR of the stellar components exceeds that of the dust component in the outskirts of the galaxy in all our sample. To establish fair comparisons, we define $r_{\rm out}$ as the radius at which the dust flux SNR drops below 1. The total flux for both the dust and stellar components (denominator in Eqs.~\ref{eq:6} and \ref{eq:7}) is then measured within this determined radius. The values of 1\,kpc and $r_{\rm out}$ both represent the physical scales in the source plane, after correcting for the lensing magnification factor $\mu$ which is calculated in Sec.~\ref{sec:lensmodel}. To ensure a reliable measurement radius and minimize potential effects from factors such as bad pixels, a requirement is imposed. Specifically, both the SNR at the boundary pixel $i$ and the subsequent pixel $(i+1)$ must be below one when determining the value of $r_{\rm out}$. We find that six of our sample galaxies have $r_{\rm out}$ smaller than 1\,kpc, indicating a low SNR ratio at the outer radii due to the small size of the galaxies. 
As a result, we exclude these six galaxies, including one $UVJ$-quiescent galaxy, from our subsequent compactness analysis. However, they are still included in the SED fitting process. Consequently, our morphology analysis includes 21 galaxies in the following discussion. Five of these unresolved galaxies have stellar mass lower than $\rm{10^{9.5}~M_\odot}$. This exclusion introduces a sample bias towards more massive galaxies, which should be taken into consideration. While our analysis does not specifically address the mass dependence, our sample may be relatively less complete in the low-mass regime.

The defined flux-based compactness parameter $c$ is a useful metric to describe the spatial distribution of flux within a galaxy.  A higher value of $c$ indicates that the galaxy has more flux concentrated in its inner region, suggesting a more compact distribution of emission.
By comparing the values of $c_{\mathrm{stellar}}$ and $c_{\mathrm{dust}}$, we can investigate the difference in spatial distribution between the PAH emission and the stellar continuum. 

Thanks to the gravitational lensing magnification effect, the size scale of our defined inner radius (i.e.,  1\,kpc) on the source plane is magnified, resulting in a larger size than the MIRI PSF size in the observed plane for all galaxies in our sample. While this inner radius is set rather arbitrarily (which is often used to infer the bulge component of galaxies at the redshift range), our scope in the discussion below is to evaluate the {\it relative} compactness of dust component over the stellar component i.e. $c_{\rm dust}/c_{\rm stellar}$. This way, our discussion is less dependent on the choice of the inner radius or outer radius definition, nor the effect of lens magnification.

\subsubsection{Non-Parametric Morphology: One-Dimensional Elliptical Flux Density Profiling}\label{sec: 1d model introduction}

Several non-parametric models are available to describe the light profile of galaxies \citep{Lotz04, Schreiber11, Baes20, Bignone20}.
We adopt one-dimensional elliptical flux density profile to describe the light distribution of our sample galaxies. The key steps are as follows:
\begin{itemize}

    \item {\bf Galaxy parameters determination}: To avoid additional assumptions, we only adopt the necessary parameters for our subsequent non-parametric morphological analysis, including pixel coordinates, axis ratio and position angle, all derived with {\tt{SExtractor}}. 
    \item {\bf Flux density calculation}: From the photo-centroid, we generate a set of concentric ellipses, starting from the innermost region and extending towards the outer region. The axis ratio and position angle of each ellipse are determined based on the {\tt{SExtractor}} output. The width of each annulus is set to one pixel.
    To calculate the flux density within each annulus, we sum up the pixel values within that region and divide by the number of pixels. This procedure is repeated for all the annuli, and the resulting flux densities are used to construct a radial profile of the galaxy's brightness. We repeat this procedure for both F200W and F770W images.
\end{itemize}

We apply the two steps above on each galaxy in the original F770W image and the PSF-matched F200W image, to extract one-dimensional flux density radial profiles, $f_{\rm F770W}(r)$ and $f_{\rm F200W}(r)$. Hereafter, we refer to this method as non-parametric, as opposed to ``parametric'' to be introduced in the following section. 

The non-parametric method offers distinct advantages by not imposing assumptions about the light profile of the galaxy, except for the axis ratio and position angle, which makes it more flexible and easier to implement. This method directly measures the flux on the original observed images, providing an intuitive and direct estimation of the galaxy's compactness. As a drawback, this method can be susceptible to errors in the outer regions of the galaxy image, the presence of fluctuations in the image background can also introduce uncertainties, particularly when assessing galaxies with low surface brightness levels.



\subsubsection{Parametric Morphology: Modeling Galaxy Structure with the S\'{e}rsic Function}\label{sec:parametric}
Parametric methods model a galaxy's light distribution using simple shapes or a combination of them. Among various parametric fitting models, the elliptical S$\acute{\mathrm{e}}$rsic model \citep[][]{1963BAAA....6...41S} is a widely adopted method for modelling the light distribution of galaxies due to its flexibility and robustness. It has become one of the most commonly used models to parameterize galaxy morphology. The S$\acute{\mathrm{e}}$rsic model is expressed as:

\begin{eqnarray}
I(r)\propto \exp \left\{ -b_n[(\frac{r}{R_{\mathrm{e}}})^{1/n}-1]\right\}
\end{eqnarray}

\noindent
where $R_\mathrm{e}$ is the effective radius, which represents the radius within which half of the total light of the galaxy is included, $n$ is the S$\acute{\mathrm{e}}$rsic index, and $b_n$ is an $n$-dependent normalization parameter ensuring that $R_\mathrm{e}$ encloses half the light. Galaxies with smaller $R_\mathrm{e}$ and higher $n$ generally exhibit more compact structures. 

 To fit the light profile of each galaxy, we utilize the software 
 {\tt{galfit}} \citep{peng2002, peng2010}. 
 For each galaxy, we generate cutouts of the F200W and F770W bands, along with their corresponding sigma images, each with dimensions of $100 \times 100$\, pixels (equivalent to $3.\!''2 \times 3.\!''2$). The magnified effective radii of our sample galaxies range from 0.\!''168 to 1.\!''48. Therefore, the cutout dimension is 2.2-19 times the galaxies’ effective radius, which is suitable for {\tt{galfit}} to determine the background level. We carefully select an unsaturated star in the same field as the PSF image and include the PSF image in every {\tt{galfit}} fit to account for image resolution limit, the software determines the optimal set of parameters for the model and produces a PSF-convolved image. This resulting image represents an ideal observation that would be obtained from the telescope in the absence of any noise or contamination.

We apply our decomposition algorithm (Sec.~\ref{sec:decomposition}) to the resulting noise-free PSF-convolved best-fit model obtained from {\tt{galfit}}, and calculate the compactness for both dust and stellar components with elliptical flux density profile method (Sec.~\ref{sec: 1d model introduction}). This method will be referred to as the ``parametric'' method in the following sections. Unlike the non-parametric method, the parametric approach models the galaxy's light distribution using a predefined profile, which can help recover missing flux at the outskirts of the galaxy. We note that, however, this method assumes an ideal S$\acute{\mathrm{e}}$rsic model in the fitting process, which may not accurately represent the real galaxy profile when they are in interaction or strongly lens-magnified, introducing additional systematic uncertainties
in the measurements. While magnification can introduce complexities such as distortion and irregular shapes, it is worth noting that our sample galaxies exhibit a moderate level of lensing with an average magnification factor of $\overline{\mu} = 1.75$ ± 0.14, indicating that the lensing effects are not highly pronounced. Additionally, these galaxies undergo limited shape distortion. As a result, the utilization of the S$\acute{\mathrm{e}}$rsic profile remains a valid and effective method for characterizing the light distribution within our sample galaxies.

\section{Results}
In this section, we present the results of our analyses. First, we compare the results derived from the non-parametric one-dimensional flux density profile and the parametric density profile of {\tt{galfit}} best-fit S$\acute{\mathrm{e}}$rsic model. We then explore the differences in morphology compactness for both stellar and dust components in order to address the region where star formation occurs. Additionally, we examine the relation between morphology compactness and other important physical properties, including stellar mass, specific star-formation rates (sSFR),and infrared-flux excess measured by 
$\mathrm{SFR_{IR}}/\mathrm{SFR_{UV}}$.
\subsection{Overview of the Dust-to-Stellar Compactness Ratio}

\begin{figure}
\plotone{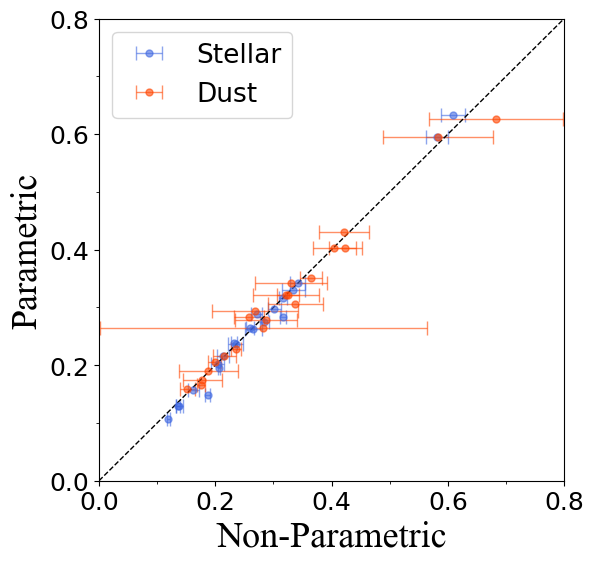}
\caption{Comparison of the flux-based compactness $c$ derived by non-parametric method (horizontal axis) and parametric method based on {\tt{galfit}} best-fit model (vertical axis). The symbol color represents dust (red) and stellar (blue) components. 
\label{fig:1d_galfit}}
\end{figure}

In Sec.~\ref{sec: 1d_profile_and_galfit}, we introduced our non-parametric and parametric methods for describing the light distribution of galaxies.  Subsequently, we determined the flux-based compactness $c$ for both the stellar and dust components using each method. The results are presented in Fig.~\ref{fig:1d_galfit}, the error bars for the non-parametric method are derived using the corresponding rms maps of each image. In contrast, the compactness measurement obtained through the parametric method relies on the best-fit model generated by {\tt{galfit}}. Consequently, there is no associated error bar for the compactness measurement based on the parametric method. Notably, we find excellent agreement between the dust and stellar components derived from both methods. Although both methods employ the same definition of compactness, they differ in their underlying assumptions and approaches, each possessing distinct strengths and limitations. The consistency in the resulting compactness estimates from these two methods implies that our approach yields robust estimates of galaxy flux concentration.

\begin{figure}
\plotone{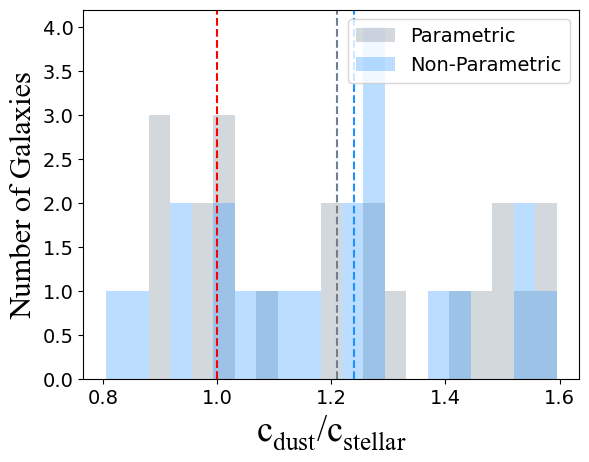}
\caption{Histogram showing the distribution of the dust-to-stellar compactness ratio of our sample (blue for the ratio calculated by the non-parametric approach and gray for the 
parametric approach). The median values derived from the two techniques (blue and gray dashed lines, respectively) are shown. The red dashed line corresponds to $\mathrm{c_{dust}}/\mathrm{c_{stellar}}$ = 1, which suggests a coherent distribution of dust and stellar. \label{fig:cd_cs}}
\end{figure}

\begin{figure*}
\plotone{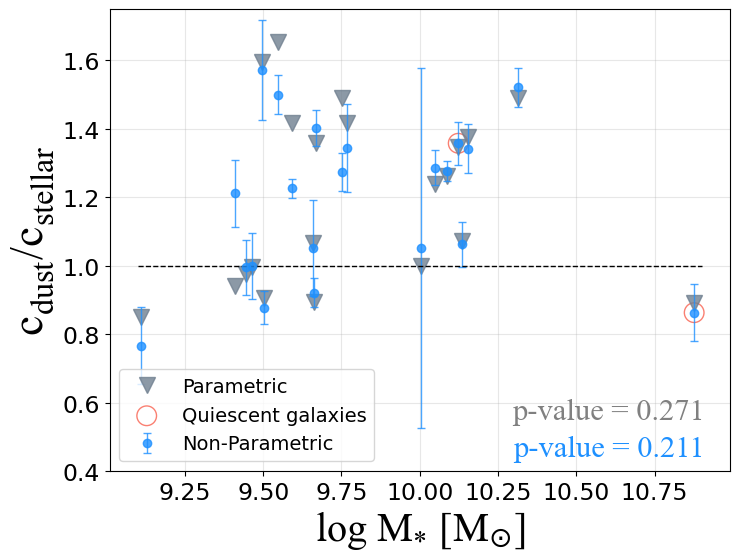}
\caption{Distribution of the dust-to-stellar compactness ratio ($\mathrm{c_{dust}}/\mathrm{c_{stellar}}$) as a function of  stellar mass ($\mathrm{M_{*}}$). 
The symbol colors represent measurements obtained from non-parametric morphology (blue) and {\tt{galfit}} parametric morphology (gray). 
The subset of $UVJ$-selected quiescent galaxies are indicated by red open circles. 
\label{fig:stellarmass}}
\end{figure*}



In Fig.~\ref{fig:cd_cs}, we present the distribution of the ratio between dust compactness, $\mathrm{c_{dust}}$, and stellar compactness, $\mathrm{c_{stellar}}$. The compactness ratio is an indicator of the central concentration of dust in a galaxy, with a higher ratio indicating more compact and dense dust emission. This measure remains unbiased by lens magnification or a specific type of the target morphology. 

We find that the majority of galaxies in our sample exhibit a dust-to-stellar compactness ratio greater than one, with 12 out of 21 galaxies exhibiting a ratio exceeding 1.1. 
The median values of the compactness ratio of the 21 galaxies in our morphology analysis are
1.24 for non-parametric method and 1.21 for parametric method.


\subsection{PAH Compactness related to Other Galaxy Physical Parameters}

In this section, we investigate the correlation between the dust-to-stellar compactness ratio and other physical parameters of galaxies. By doing so, we aim to study the physical processes responsible for galaxy structures, e.g., the growth of bulges. 

\subsubsection{Stellar Mass}
\label{sec: stellarmass}

Fig.~\ref{fig:stellarmass} shows the $\mathrm{c_{dust}}/\mathrm{c_{stellar}}$ as a function of stellar mass.  
To examine the association between stellar mass and $\mathrm{c_{dust}}/\mathrm{c_{stellar}}$, we employ the Spearman rank correlation method and calculate the p-value to evaluate the strength of this relationship. A p-value close to zero indicates that the variables are likely related and the correlation is not just a random occurrence.
Our analysis does not reveal a significant correlation between dust compactness ratio and stellar mass. Spearman rank correlation tests confirm this lack of trend, with a p-value of 0.21 for the non-parametric method and 0.27 for the parametric one. 
We note, however, that the most massive quiescent galaxy displays a less compact core with lower dust concentration, in contrast to the other less massive quiescent galaxies which exhibit a higher concentration of dust.
\subsubsection{PAH Emission Compactness and Specific Star Formation Rate}
\label{sec: ssfr}
Many previous works \citep[e.g.,][]{Allamandola1985, Allamandola1989, Tielens08, Micelotta10, Stierwalt14, Riechers14} have highlighted the crucial role of PAHs in the formation of IR emission bands seen in the spectra of star-forming galaxies. 
These molecules play an important role in cooling the gas in the ISM, facilitating the formation of compact and dense clouds that can later collapse and give rise to new stars. In order to better understand the relatively higher dust compactness observed in our galaxies relative to their stellar components, we explore the relation between the dust-to-stellar compactness ratio and sSFR. The choice of sSFR over SFR allows us to avoid any potential impact from the wide range of stellar masses in our sample on our conclusions. 


Fig.~\ref{fig:ssfr} presents our results, where sSFR is calculated by $\mathrm{SFR_{UV+IR}}/\mathrm{M_{*}}$. The calculation process for SFR\textsubscript{IR} and SFR\textsubscript{UV} are described in Sec.~\ref{sec:uvj_sfr}.
The Spearman rank correlation is employed to determine whether there is an association between the sSFR and the dust compactness ratio. 
The computed p-values for the correlation between the non-parametric method-derived compactness and the parametric method-derived compactness are 0.94 and 0.44, respectively, both indicating a lack of statistically significant correlation between these two physical properties. 

To further explore the relation, we divide the sample into three groups, each consisting of seven galaxies, based on their sSFR. Median sSFR values are then calculated for the low, medium, and high sSFR bins. The results showed that the median compactness ratio of the high-sSFR sub-sample is higher than that of the medium-sSFR sub-sample, taking the error bar into account, while being comparable to that of the low-sSFR sub-sample.


\subsubsection{PAH Emission Compactness and \texorpdfstring{$\mathrm{SFR_{IR}/SFR_{UV}}$}{SFR\_IR/SFR\_UV}}

Lastly, correlation between the compactness ratio c and SFR\textsubscript{IR}/SFR\textsubscript{UV} is investigated. SFR\textsubscript{IR}/SFR\textsubscript{UV} is an indicator of a galaxy's dustiness, with a higher value referring to a higher obscured star formation fraction. This ratio has been widely used as a tracer of dust content in galaxies, especially for those that are actively forming stars \citep{Wuyts11, Whitaker18, Inami22_alma}.

\begin{figure*}
\includegraphics[width=0.45\textwidth]{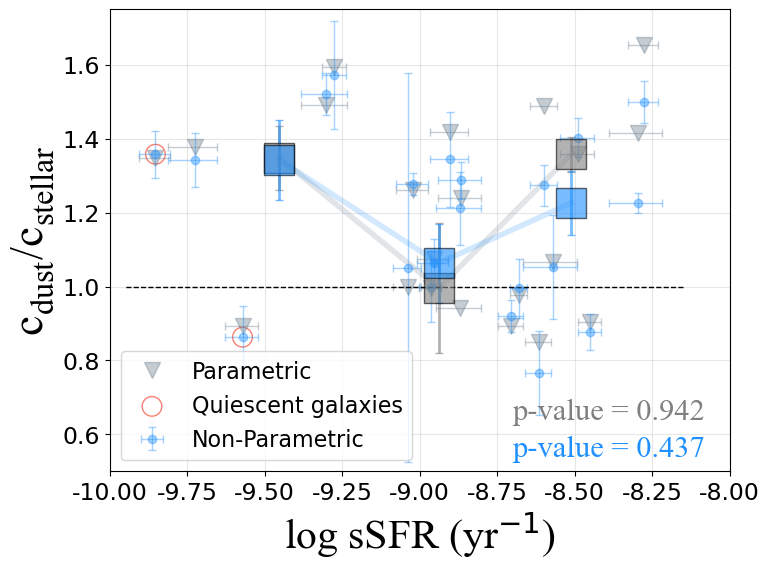}
\includegraphics[width=0.45\textwidth]{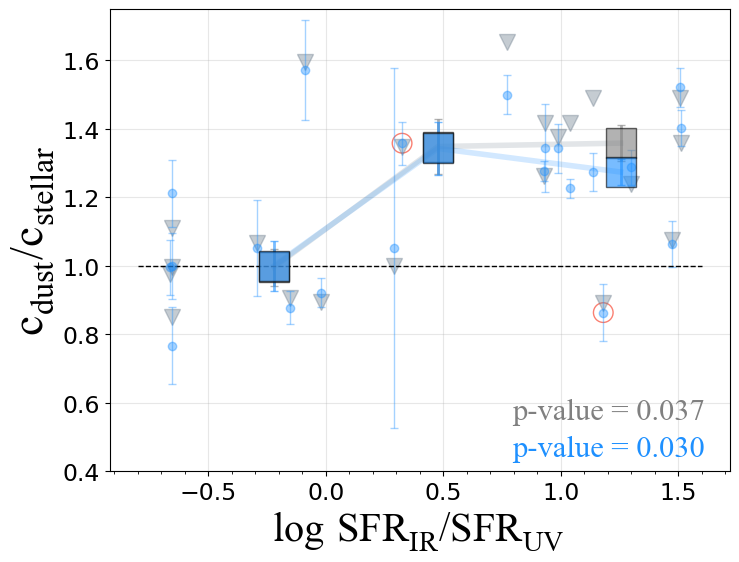}
\caption{
Same as Fig.~\ref{fig:stellarmass} but as a function of sSFR ({\it left}) and $\mathrm{SFR_{IR}}/\mathrm{SFR_{UV}}$ ({\it right}). The error bars for sSFR are computed based on {\tt{CIGALE}}'s Bayesian probability distribution. The large blue and gray square markers with error bars denote the median values and associated uncertainties (calculated using the normalized median absolute deviation with $N-1$ degrees of freedom) in each bin.
\label{fig:ssfr}}
\end{figure*}

The right panel of Fig.~\ref{fig:ssfr} illustrates the results obtained, where the calculations for $\mathrm{SFR_{IR}}/ \mathrm{SFR_{UV}}$ are same as aforementioned. To further investigate, we employ Spearman rank correlation to examine the relationship between the dust-to-stellar compactness ratio and the UV-to-IR SFR ratio. The p-values obtained from this analysis are both below 0.05, as shown in the right panel of Fig.~\ref{fig:ssfr}, indicating a strong statistical significance in the correlation between $\mathrm{SFR_{IR}}/ \mathrm{SFR_{UV}}$ and $\mathrm{c_{dust}}/\mathrm{c_{stellar}}$. We implement a similar binning strategy as in Sec.~\ref{sec: ssfr} and calculate the median values and their corresponding errors for the low, medium, and high $\mathrm{SFR_{IR}}/\mathrm{SFR_{UV}}$ bins.

\section{Discussions}
\label{sec: discussion}
\subsection{Investigating the Correlation between Dust Morphology and Physical Properties}

In Sec.~\ref{sec: stellarmass}, we investigate potential correlation between the dust-to-stellar compactness ratio and stellar mass. 
As our compactness ratio is defined by fluxes within the central 1\,kpc, our findings is likely attributed to the growth of bulges in the central regions. We find a diverse range of dust compactness among galaxies with a stellar mass between $\rm{10^{9.5}<M_*/M_\odot<10^{10.5}}$, some of which reaches $\sim1.6$. This result aligns with the notion that rapid bulge growth through active star formation occurs in the central regions of galaxies in this mass range at this cosmic time \citep[e.g.,][]{Bell12, tadaki17, Gargiulo2019}. 
The on-set of bulge formation is of particular interest \citep{Larson80, Kennicutt94, Lucia06}. During cosmic noon, a significant proportion of galaxies exhibit the emergence of a dense core at ${\rm M_* \sim 10^{10}\,M_\odot}$ and beyond \citep[e.g.,][]{barro17}. The emergence of a dense central core is attributed to the formation of a bulge component through mergers and gas infall, or `compaction' \citep{Zolotov15, barro16, Tacchella18, Lapiner23}. The observed increase of dust-compact galaxies at $\gtrsim 10^{9.5}\,M_\odot$ in our sample thus likely reflect galaxies in the bulge-formation phase.

It is still an open question if a large fraction of galaxies lower than the mass range ($\rm{M_*<10^{9.5}~M_\odot}$) explored here have a compact dust core or not. In our sample, even with the help of gravitational lens magnification, five out of eight galaxies in the mass range are not resolved due to their intrinsically small size 
(see Appendix.~\ref{fig:stamps_unresolve}). Future deep observations in lensing fields, where strong(er) magnification helps resolving such small sources, will be critical for further statistical arguments on the onset of bulge formation in galaxies.

As discussed in Sec.~\ref{sec:uvj_sfr}, we identify three galaxies are located in the quiescent region of the $UVJ$ diagram but near the boundary to the star-forming region. From the location of these galaxies in the $UVJ$ diagram, supported as well by their location below the star-forming main sequence (Fig.~\ref{fig:main sequence and uvj}), we interpret these galaxies as dusty post-starburst galaxies that have recently quenched their star-forming activities, which makes them detectable in F770W. Among the $UVJ$-quiescent galaxies in our sample, one galaxy satisfies the post-starburst classification according to \cite{Belli19}, while all the $UVJ$-quiescent galaxies fall within the modified post-starburst region defined in \cite{Mao22}.

We also identify a potential outlier, ID~194, the most massive galaxy in our sample. Unlike most of the less massive galaxies, it has a lower dust compactness ratio with $\mathrm{c_{dust}}/\mathrm{c_{stellar}}$ = 0.86 (non-parametric)/ 0.89 (parametric), indicating a less compact dust structure compared to its stellar structure. The postage stamps 
also clearly show that ID~194 has a bright and compact stellar core as seen in F200W, with a relatively extended dust structure as revealed in F770W and F1000W (Appendix.~\ref{fig:stamps1}). 
This galaxy is classified as $UVJ$-quiescent and thus is relatively evolved (while its detection in MIRI indicates a little ongoing star formation, which is consistent with its location on the mass-SFR plane; Fig.~\ref{fig:main sequence and uvj}). Compact stellar cores in quiescent galaxies of this mass range have been reported in previous studies \citep{van09, szomoru12, morishita14, abd23}. One potential explanation is the inside-out growth and subsequent inside-out quenching process, whereby star formation activities move outward as galaxies age, or through mergers in quiescent galaxies \citep{Schreiber09, Schreiber18, law09, nelson12, nelson13, wilman20, Matharu22}. 

In a recent study, \cite{shen23} used MIRI data from the Cosmic Evolution Early Release Science survey (CEERS; \citealt{Finkelstein23}) to investigate the morphologies of star-forming galaxies in the rest-MIR and trace obscured star formation regions. Their findings indicate that the effective radius measured in the rest-MIR band is slightly smaller than that in the rest-optical band, which traces the stellar continuum. This result is consistent with our own conclusions.
 While their sample includes galaxies up to $\sim 10^{11} M_\odot$, they did not report the lower dust concentration that we observed in our most massive galaxy. The discrepancy observed between our work and the findings of \cite{shen23} may be attributed to the divergent focus of their study, which specifically examined star-forming galaxies.
 Additionally, our approach offers a unique advantage in that we decompose dust component from observed F770W profile, allowing for a direct comparison of the light distributions between the stellar continuum and PAH emission, instead of simply comparing observed light profile in two bands. The use of both parametric and non-parametric methods further enables us to avoid potential bias from the model. 
Moreover, our study stands out for the use of background galaxies from the lensing cluster SMACS0723, which enables us to study magnified galaxies with reduced impact from the PSF.

A recent study by \cite{Matharu23}, using spatially resolved Balmer Decrement observations via NIRISS, reported a discovery of flat dust attenuation profiles in massive galaxies ($\rm{10^{10}<M_*/M_\odot<10^{11.1}}$), while they found more centrally concentrated profiles observed in galaxies within a stellar mass range of $\rm{10^{7.6}<M_*/M_\odot<10^{10}}$. This is in full agreement with our findings and interpretation of the observed stellar mass trend of the compactness ratio.

In addition to stellar mass, we also investigate the relation between dust compactness and sSFR in Sec.~\ref{sec: ssfr}. Although our analysis does not reveal a statistically significant correlation between sSFR and dust compactness, it should be noted that the majority of our sample galaxies are more or less typical star-forming galaxies i.e. those on/near the star-forming main sequence (Figs.~\ref{fig:main sequence and uvj} and \ref{fig:ssfr}). In fact, we find that the median compactness ratio is higher in the high sSFR bin than in the medium sSFR bin, while remaining comparable to that of the low sSFR bin. 
However, high sSFR galaxies are known to have strong radiation fields, which can lead to the destruction of PAH molecules through photodissociation and other mechanisms \citep{Voit92, Siebenmorgen04, Sales2010, 
Micelotta10_feb, Monfredini19}. The most massive galaxy in our sample, with significantly decreased compactness ratio compared to other galaxies, may be one that experienced significant destruction of dust grains during its bulge growth and/or quenching phases. Therefore, the correlation between sSFR and PAH abundance could be degenerate, which provides a plausible explanation for the observed lack of correlation.

To summarize, the higher compactness ratio traced by flux inside 1\,kpc provides critical insight into the relationship between bulge growth and dust concentration in galaxies. However, the observed trend is limited by the sample size analyzed in this study. To better understand the formation scenario of galaxy bulges, it would be valuable to expand the sample to include a wider range of stellar masses and sSFR. Such a study will allow further characterization of the correlation between dust compactness ratio and galaxy properties and help us shed light on the formation of galaxy bulges in more detail.

\subsection{Effective Use of MIRI to Study Galaxy Structures}

In this study, we investigate the morphologies of PAH-emitting galaxies utilizing the F200W and F770W filters. By selecting galaxies in a specific redshift range, we are able to effectively trace existing stellar components with F200W, while also tracing ongoing star-forming activity inside galaxies using the decomposed 3.3 $\mu$m PAH feature with F770W. Our analysis, based on the flux-based compactness parameters $c_\mathrm{stellar}$ and $c_\mathrm{dust}$, reveals that the majority of our sample exhibited a compact dust core compared to the stellar component, which suggests more enhanced star-forming activity in the central regions of these galaxies. 
This result is consistent with previous studies that also reported centrally concentrated dust emission predominantly relied on FIR ALMA observations \citep{simpson15, hodge16, barro16, tadaki17, tadakiapj17, elbaz18, Calistro18, fujimoto18,lang19, tadaki20, chen20, cheng20, ikeda22, guijarro22}. FIR emissions are mainly attributed to dust continuum and are commonly used as a tracer of star formation processes, including intense starburst activity in the central regions of galaxies, extended star formation in galaxy disks, and bursts of star formation triggered by galaxy mergers \citep{kennicutt98, Calzetti2000, Casey2014}. However, FIR emission may be biased towards more active star-forming galaxies, including those with extreme starburst activity, and can be affected by other physical processes, such as AGN activity and dust heating by evolved stars, with a typical timescale of $\sim100$\,Myr \citep{kennicutt98_2, Elbaz11}. Besides, those studies using ALMA observations were limited to more massive galaxies with $M_*/~M_\odot > 10^{10.5}$ or stacking analysis, posing challenges in achieving a comprehensive understanding of the evolution of galaxy populations across a broader range of 
stellar masses.

Similarly, the use of spatially resolved \ha\ maps for a large sample has been enabled by slitless spectroscopy 
\citep{nelson16,nelson16_balmer, Vulcani16, wilman20, Matharu22}. Massive O and B stars produce \ha\ emission through ionizing photons, which have a lifetime of a few million years, implying that \ha\ emission is a tracer of more recent star formation activity (occurring in a few million years, \citealt{kennicutt98_2}). Additionally, it was reported by \citet{nelson16} that more massive galaxies tend to exhibit more extended \ha\ emission, while the size of the \ha\ region and the stellar continuum region is rather consistent for the low mass range, indicating ongoing star-formation activity at larger radii than existing stars, corroborating the inside-out growth theory. Furthermore, \cite{nelson18} conducted a study comparing the sizes of \ha\, dust, and UV components in a progenitor of an Andromeda galaxy at redshift $z=1.25$. Their analysis revealed more compact dust profiles compared to the sizes of both \ha\ and UV regions. 
However, it is important to consider the influence of dust extinction on \ha\ measurements. Local star-forming galaxies typically experience an average dust extinction of approximately 1 magnitude, while at redshifts $z = 1-2$, this value can range from 1 to 3 magnitudes \citep{Koyama10}. As discussed in \cite{nelson16_balmer} and \cite{Tacchella18}, the dust attenuation is highest and most concentrated towards the \ha\ emission in more massive galaxies. This effect might be more significant in the inner region of the galaxies, where vigorous star formation occurs, and a large amount of dust and gas exists.

High-resolution MIRI imaging offers a solution to those limitations and can complement ALMA and \ha\ observations. First of all, there is no need for pre-selection of targets as required for ALMA due to its narrow spectral window. This helps us construct a more complete sample. Moreover, compared to \ha, mid-IR is less affected by dust extinction due to its longer wavelength. Therefore, with MIRI, we are able to access a more accurate picture of star-forming activities within galaxies. These advantages make our method a comprehensive and robust approach to study galaxy morphology. 

\section{Summary and Future Prospect}
\label{sec: summary}
This paper presents an innovative approach for investigating the morphological characteristics of galaxies in the JWST ERO field SMACS0723, utilizing mid-IR imaging with the MIRI instrument.
The key findings of this investigation are summarized below:
\begin{itemize}
    \item Our study is based on the analysis of galaxies detected through the MIRI F770W filter. We meticulously conduct a cross-matching process between the MIRI F770W galaxy sample and the NIRCam+HST catalog, resulting in a sample of 27 sources for structural analysis.
    
    \item We utilize the SED fitting tool {\tt{CIGALE}} with flux corrected for lensing magnification to derive the intrinsic physical properties of our sample galaxies. Our analysis shows that the majority of the galaxies in our sample are main sequence star-forming galaxies. However, we also observe three galaxies located near the boundary that separates the star-forming and quiescent regions.
    
    \item We develop an algorithm to decompose both stellar and dust components by combining F200W and F770W images and the SED results. To study the flux distribution of both components, we introduce a flux-based compactness parameter, represented by the ratio of the flux within 1\,kpc to the total flux, denoted as $c$. We use this parameter to quantify the flux compactness of a sub-sample of 21 resolved galaxies. We calculate $c$ by employing both parametric and non-parametric techniques. The resulting compactness measurements from both approaches exhibit excellent agreement, indicating the effectiveness of our proposed method for measuring morphology compactness of our sample galaxies.

    \item We find variations in dust-to-stellar compactness ($c_{\rm dust}/c_{\rm stellar}$) across galaxies with stellar masses $\rm{10^{9.5}<M_*/M_\odot<10^{10.5}}$. Using non-parametric and parametric methods, we observe average $c_{\rm dust}/c_{\rm stellar}$ for these galaxies of 1.23 (±0.05) and 1.27 (±0.06), respectively. Furthermore, the most massive galaxy in our sample exhibits a comparable level of compactness between its stellar core and dust, with a dust-to-stellar ratio of 0.86 (0.89) as determined using non-parametric (parametric) methods, respectively. This suggests that the intense star-forming activities in the core region may have already
    undergone quenching. We also note a lack of correlation between dust-to-stellar compactness and sSFR. While the highest sSFR bin exhibits a higher median value for the compactness ratio compared to the medium bin, the overall trend remains inconclusive. This lack of correlation could potentially arise from the degeneracy between sSFR and PAH emission. We further investigate $\mathrm{SFR_{IR}}/\mathrm{SFR_{UV}}$, a well-established tracer of dust extinction in galaxies. Our analysis reveals a positive correlation between the $\mathrm{SFR_{IR}}/\mathrm{SFR_{UV}}$ ratio and $c_{\rm dust}/c_{\rm stellar}$. This indicates that galaxies with a higher $\mathrm{SFR_{IR}}/\mathrm{SFR_{UV}}$ ratio may exhibit more intense or concentrated star-forming regions.
\end{itemize}

This study serves as a pilot investigation into the morphologies of galaxies using JWST MIRI observations and demonstrates the potential of JWST in examining galaxy structure. A future joint analysis of multiple Cycle 1 program will offer a  larger sample and a more complete understanding of the structure formation of stellar and ISM components in galaxies across the cosmic time. 

\begin{acknowledgments}

We would like to express our gratitude to the anonymous referee and editor for their insightful comments that enhanced the manuscript. 
We acknowledge ERO production team who developed, executed and compiled the ERO observations. The data used in this work were made available through MAST at Space Telescope Science Institute.
ZL would like to thank Drs. Jose Manuel P\'erez-Mart\'inez, Rieko Momose and Lilan Yang for the valuable discussion that improved our analysis. 
ZL acknowledges support from JST SPRING, Grant Number JPMJSP2114 and Kakenhi International Leading Research (\#22K21349).
Support for this study was provided by NASA through a grant HST-GO-15804 from the Space Telescope Science Institute, which is operated by the Association of Universities for Research in Astronomy, Inc., under NASA contract NAS 5-26555. \end{acknowledgments}



\software{\texttt{Astropy} \citep{Astropy13,Astropy18},
          \texttt{SExtractor} \citep{bertin1996},
          \texttt{CIGALE} \citep{2019A&A...622A.103B}}



\appendix
\label{appendix}

\section{The Cutout Images For Each Galaxy}\label{sec:appA}
\begin{figure}
  \centering
  \includegraphics[height=0.93\textheight]
  {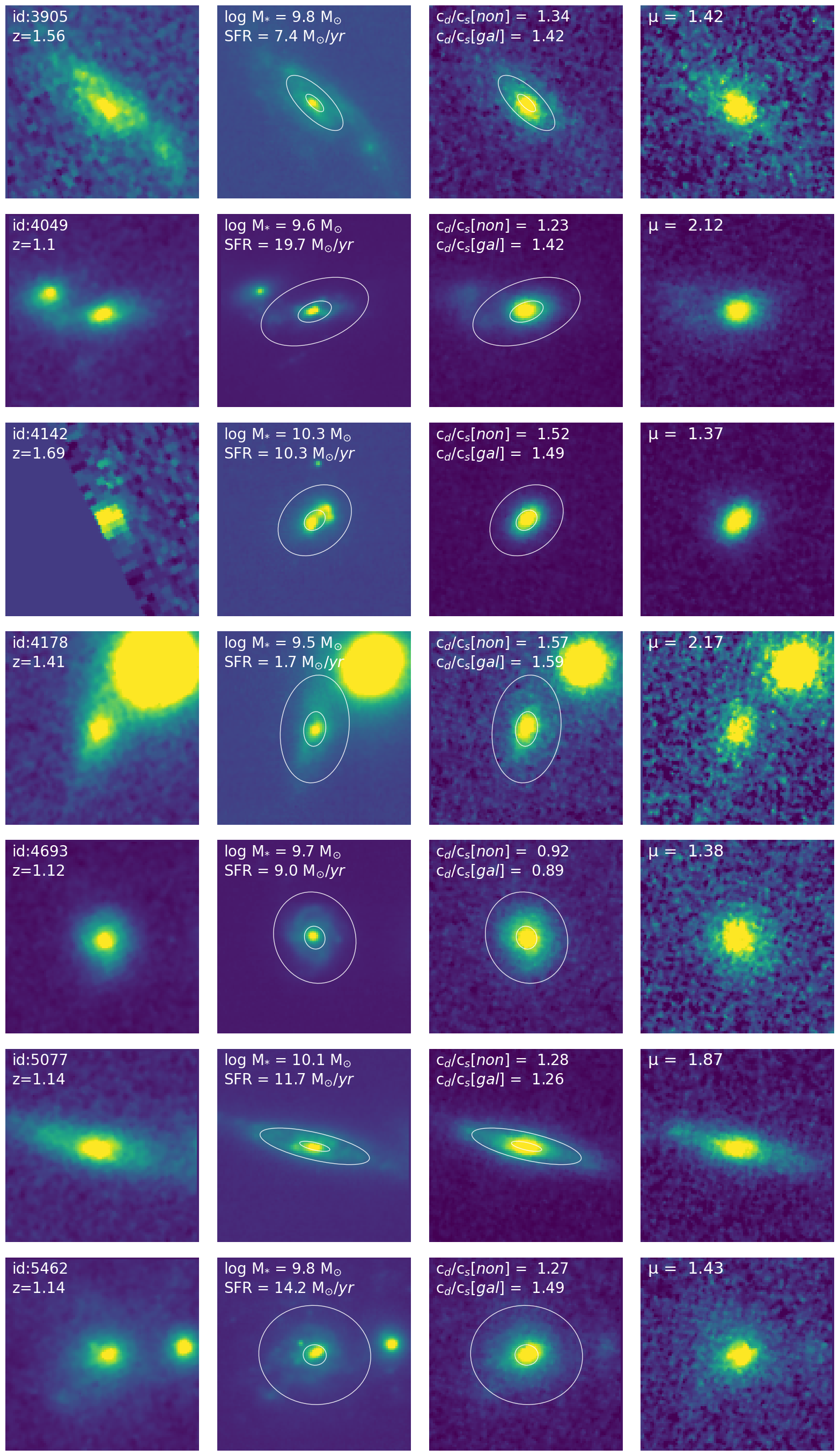}
  \caption{Postage stamps of the selected galaxies used for compactness analysis. Each column shows the galaxy centered in the F160W, F200W, F770W, and F1000W filters from left to right, with a size of $3.!''2 \times 3.!''2$. 
  Stellar mass and SFR calculations are detailed in Sec. \ref{sec:stellarmass_sed}, with both values corrected for the lensing magnification. The two white ellipses in the middle columns represent $r_{\rm in}$ and $r_{\rm out}$, which are used for calculating compactness. The lensing magnification factor ($\mu$) for each galaxy, as calculated in Sec.~\ref{sec:lensmodel}, is also provided.}  
  \label{fig:stamps3}
\end{figure}

\begin{figure}
  \centering
  \includegraphics[height=0.857\textheight]{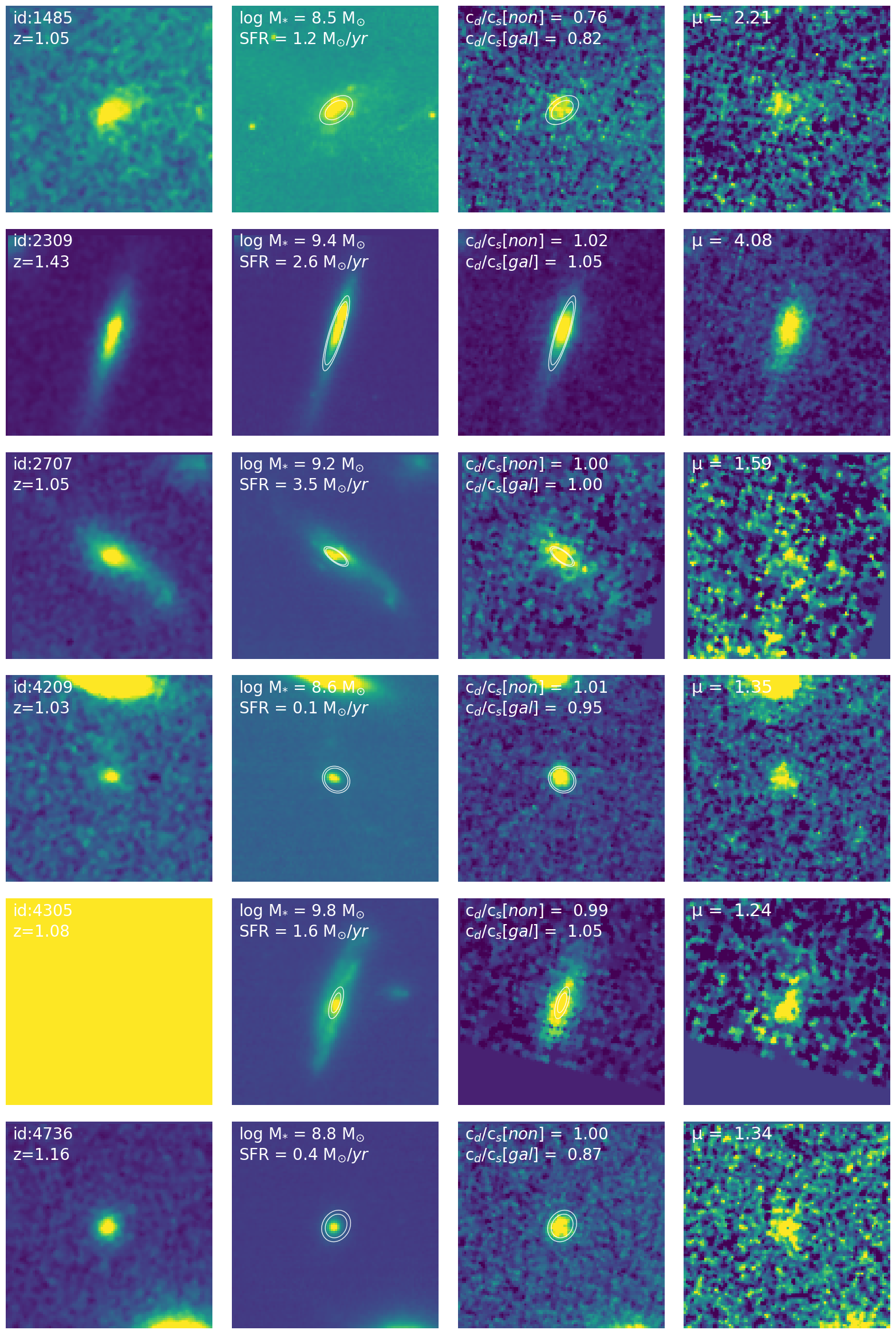}
  \caption{Postage stamps of galaxies excluded from the morphology analysis due to their small size and low mass, but included in the physical properties analysis. For the exclusion criteria for unresolved galaxies, see Sec.~\ref{sec: 1d model introduction}.}
   \label{fig:stamps_unresolve}
\end{figure}

{Fig.~\ref{fig:stamps3} displays postage stamps of our resolved galaxies in HST WFC3 F160W, JWST NIRCam F200W, JWST MIRI F770W, and F1000W, along with the corresponding inner and outer radii. On the other hand, Fig.~\ref{fig:stamps_unresolve} shows postage stamps of six unresolved galaxies in the same bands. It is worth noting that Galaxy 4305 in Fig.~\ref{fig:stamps_unresolve} was not observed by HST, and was included in the physical properties analysis despite there being no coverage in the southern region in F770W and F1000W. To ensure that the flux measurement is not biased due to this lack of coverage, we inspect the Kron radius within which the flux is counted, and we find that the Kron radius does not extend to the region without coverage. Therefore, the flux-based SED fitting is not affected by the absence of coverage in that particular area.}

\begin{figure*}
  \centering
  \includegraphics[height=0.93\textheight]
  {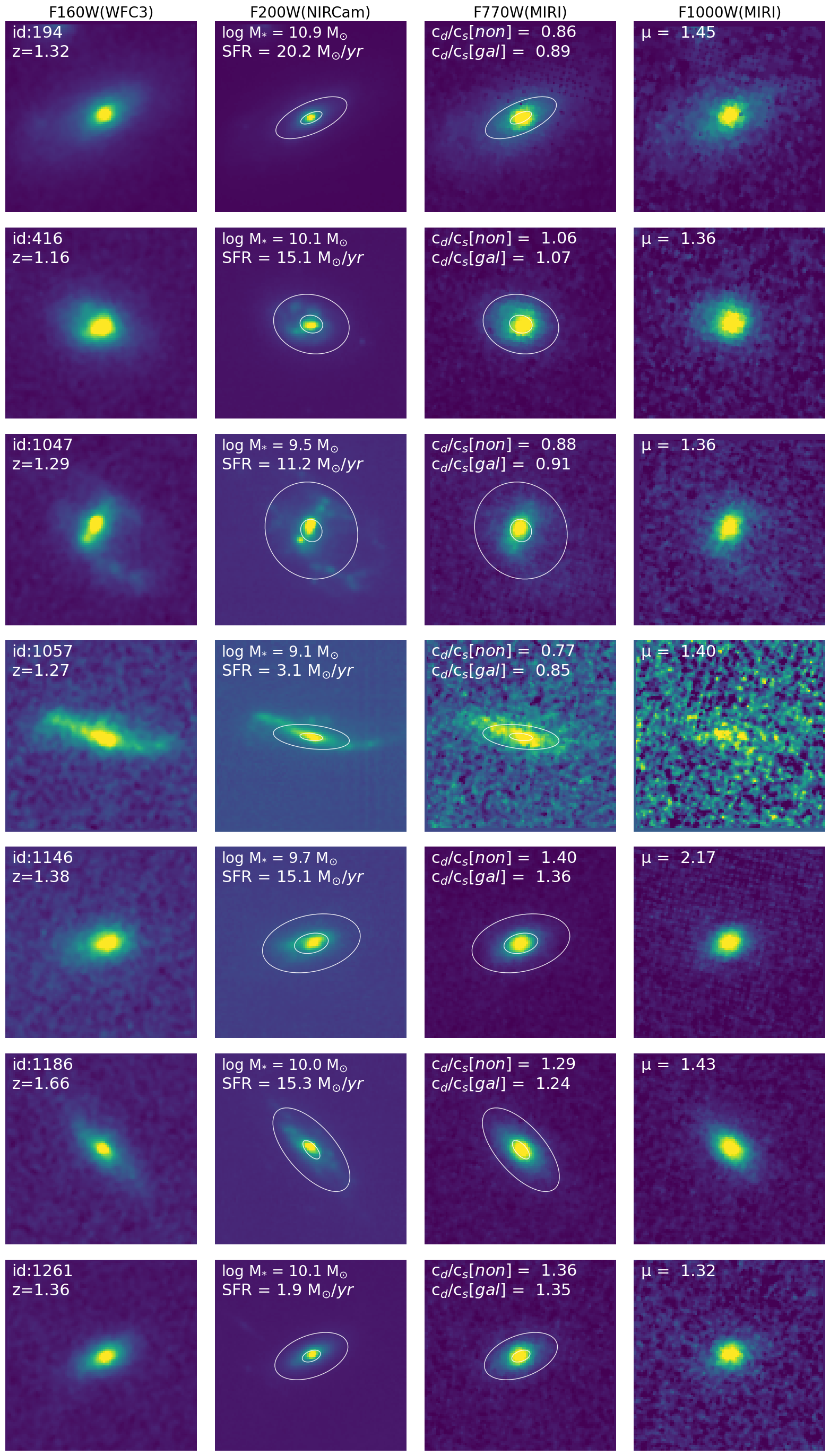}

  \label{fig:stamps1}
\end{figure*}

\begin{figure*}
  \centering
  \includegraphics[height=0.93\textheight]
  {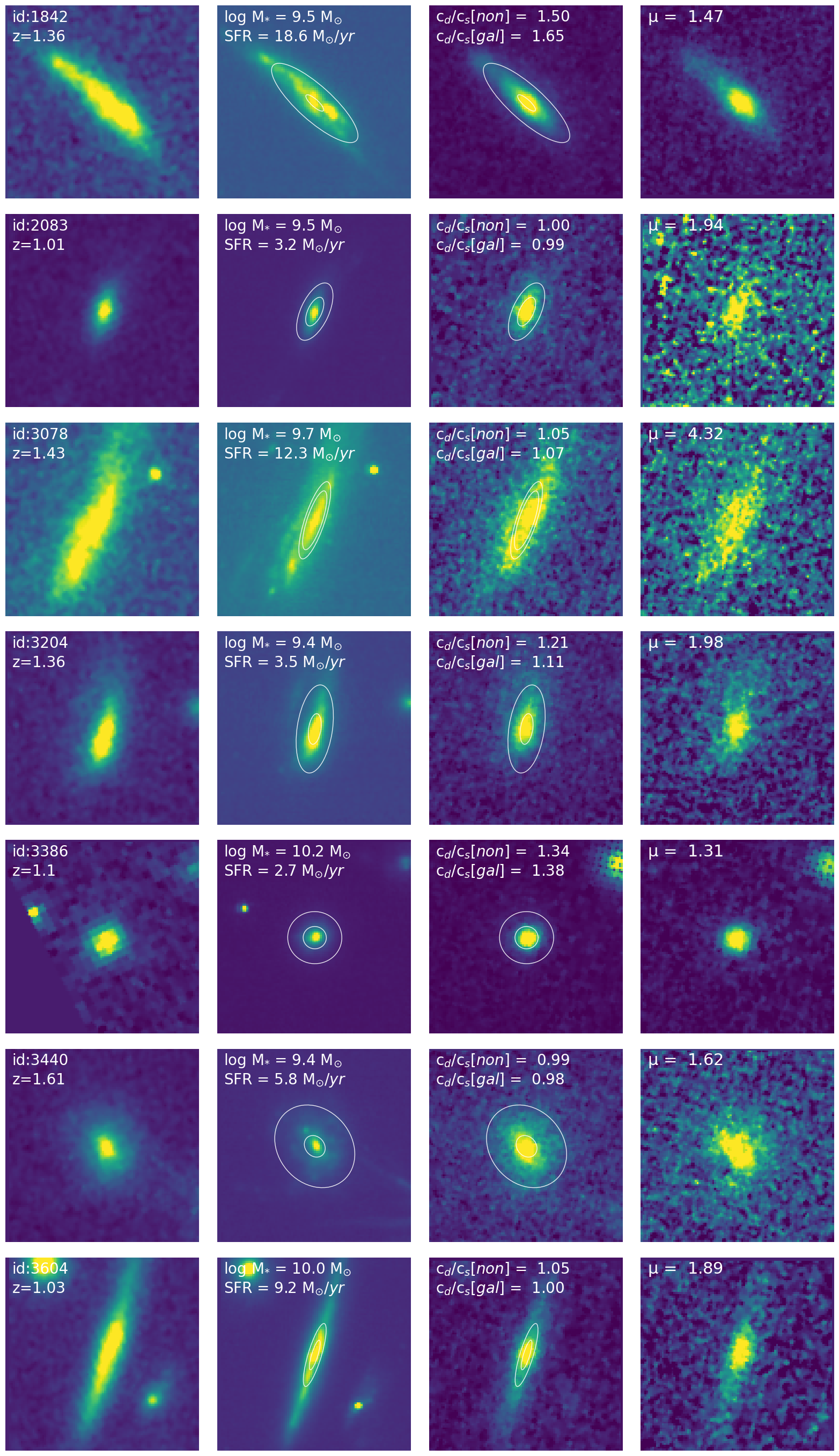}
  \label{fig:stamps2}
\end{figure*}

\bibliography{sample631}{}
\bibliographystyle{aasjournal}



\end{document}